\documentclass [aps,prb,twocolumn,groupedaddress,showpacs] {revtex4}

\usepackage {graphicx,amssymb,amsbsy,color}
\begin {document}

\title {
Localized states in the continuum in low-dimensional systems
}
\author {Khee-Kyun Voo$^*$ and C. S. Chu}
\affiliation {Department of Electrophysics, National Chiao Tung
University, Hsinchu 30010, Taiwan, Republic of China}

\date {\today}

\begin {abstract}

It is shown in this paper that for open systems, states which are 
localized in space, discrete in energy, and embedded in the continuum of
extended states, can be sustained by low-dimensional and channeled leads.
These states have an origin different from that of analogous states
discussed by J. von Neumann and E. Wigner [Phys. Z. {\bf 30}, 465 (1929)].
A few representative systems are discussed. These states cause, for
example, infinitely sharp Fano resonance in transport when they are
marginally destroyed.

\end {abstract}

\pacs {73.22.Dj, 03.65.Ge, 73.23.Ad, 73.63.-b}
\maketitle

\newpage

\section {Introduction}

Shortly after the discovery of quantum mechanics, von Neumann and Wigner
pointed \cite {NW29,SH75} out that potentials defining closed
boundary conditions were not the only cause of discrete and localized 
(normalizable) states. They pointed out that localized states could also
be due to the destructive interference in the Bragg scattering from
certain long-ranged wiggling potentials defining open boundary conditions.
These states are embedded in the continuum, decay in space with a power
dependence, and have been studied in atomic and molecular
systems \cite {SH75,FW85,CFR03} and superlattices. \cite {CSF92} In this
paper, we show that analogous states can also be found in open systems
with low-dimensional leads. They decay exponentially in space in contrast
to those discussed by von Neumann and Wigner, \cite {NW29,SH75} and also
have a different origin. Moreover, they are shown to be related to the
infinitely sharp Fano resonance \cite {Fan35} in transport.

To illustrate the properties of such states in low-dimensional systems,
three representative model systems will be discussed --- A tight-binding
(TB) molecular system, a quantum graph with doubly-connected
one-dimensional (1D) channels, and a waveguide in a two-dimensional (2D)
space. All three are open systems. The first two are simple enough for
analytic analyses, where some generic properties can be studied
rigorously. The third one is to illustrate the presence of these states
in more realistic systems, but unfortunately, it allows only a numerical
analysis.

\section {Models and Discussions}

\subsection {A tight-binding molecular system}
\label {tbmol}

\begin {figure}


\includegraphics{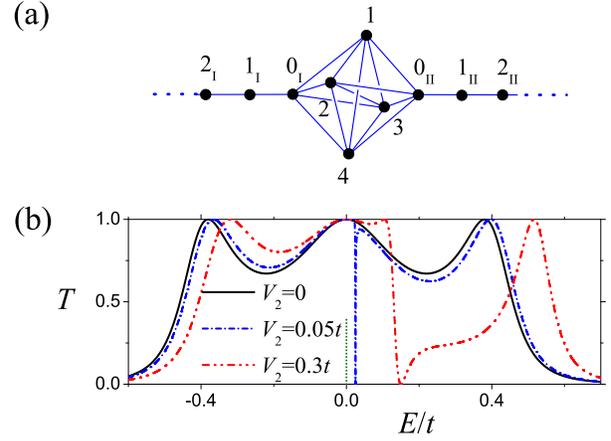}
\vspace {7cm}

\caption {
(Color online) 
(a) The considered open TB system, which consists of a molecule with four
sites (labeled by 1 $\sim$ 4) connected to two leads (labeled by I and II)
with serially connected sites (labeled by $i_\eta$ for lead $\eta$, where
$i= 0 \sim \infty$ and $\eta=$ I and II). The hoppings are denoted by
bonds.
(b) The transmission probability $T$ at different energies for $V_2=0$
(solid line), where there is a LSC (located by a dotted line); and $V_2 = 
0.05 t$ (dash-dotted line) and $0.3 t$ (dash-dot-dotted line), where
there are no LSC. Those not-mentioned system parameters are refered to the
text.
}

\label {mol}
\end {figure}

First we consider the TB system shown in Fig.~\ref {mol}(a) which is 
defined by the Hamiltonian
\begin {eqnarray}
H &=& H_{\rm mol} + H_{\rm lead} + H_{\rm mol-lead},
\nonumber\\
H_{\rm mol} &\equiv&  \sum_{i=1}^4 c_i^\dagger V_i^{~} c_i^{~} +
\sum_{i=1}^4 \sum_{j>i} \left[ c_i^\dagger h_{ij}^{~} c_j^{~} + {\rm H.c.}
\right], \nonumber\\
H_{\rm lead} &\equiv&  - t \sum_{\eta = {\rm I, II}}  
\sum_{i=0} ^\infty 
c_{(i+1)_\eta}^\dagger c_{i_\eta}^{~} + {\rm H.c.}, \nonumber\\ 
H_{\rm mol-lead} &\equiv& \sum_{\eta = {\rm I, II}} \sum_{i=1}^4
c_{0_\eta} ^\dagger h_{0_\eta i}^{~} c_i^{~} + {\rm H.c.}, 
\end {eqnarray}
where $c_\lambda$, $\lambda \in \{ 1 \sim 4, 0_{\rm I} \sim \infty_{\rm
I}, 0_{\rm II} \sim \infty_{\rm II} \}$, is the annihilation operator of a
spinless particle on site $\lambda$, $t$ and $V_i$ are real, and $h_{ij}$
and $h_{0_\eta i}$ are complex in general. Sites $1 \sim 4$ are in a
``molecule'', and sites $0_\eta \sim \infty_\eta$ are in lead $\eta$,
$\eta=$ I and II. This Hamiltonian may describe a nanoscopic molecule or a
mesoscopic cluster of quantum dots connected to two leads.

Using a basis set $\{ | \lambda \rangle \}$ defined by $| \lambda
\rangle \equiv c_\lambda ^\dagger | 0 \rangle$ and $c_\lambda | 0 \rangle
\equiv 0$, one can write the time-independent Schr$\ddot {\rm o}$dinger
equation (TISE) $H | \psi \rangle = E | \psi \rangle$, where $E$ is the
energy of the particle, into a set of simultaneous finite-difference (FD)
equations. In a lead, the FD equations read $t \psi_{j-1}^\eta + E \psi_j
^\eta + t \psi_{j+1}^\eta = 0$, for $j=1 \sim \infty$ and $\eta = $ I or
II, where $\psi _{j} ^\eta \equiv \langle j_\eta | \psi \rangle$. It has
an analytic solution 
\begin {eqnarray}
\psi_j^\eta = A_\eta e^{-ikj} + B_\eta e^{ikj},
\label {wflead}
\end {eqnarray}
where $A_\eta$ and $B_\eta$ are arbitrary complex numbers, and $k \equiv 
|k| \equiv {\rm cos}^{-1} [-E/(2t)] \equiv k (E)$. Since $k>0$, the
ingoing or inward propagating wave amplitudes are $A_{\rm I}$ and $A_{\rm
II}$.

There remains six FD equations not solved by Eq.~(\ref {wflead}). 
Replacing $\psi_0^\eta$ and $\psi_1^\eta$ by $A_\eta$ and $B_\eta$ using
Eq.~(\ref {wflead}), and writing $\psi _{j} \equiv \langle j | \psi
\rangle$ for $j = 1 \sim 4$, the six equations can be written as a matrix
equation,
\begin {widetext}
\begin {eqnarray}
\left[
\begin {array} {cccccc}
- t e^{ik (E)} - E & h_{0_{\rm I} 1}^{~} & h_{0_{\rm I} 2}^{~} & h_{0_{\rm
I} 3}^{~} & h_{0_{\rm I} 4}^{~} & 0 \\
h_{0_{\rm I} 1}^* & V_1-E & h_{12}^{~} & h_{13}^{~} & h_{14}^{~} &
h_{0_{\rm II} 1}^* \\
h_{0_{\rm I} 2}^* & h_{12}^* & V_2-E & h_{23}^{~} & h_{24}^{~} & h_{0_{\rm
II} 2}^* \\
h_{0_{\rm I} 3}^* & h_{13}^* & h_{23}^* & V_3-E & h_{34}^{~} & h_{0_{\rm
II} 3}^* \\
h_{0_{\rm I} 4}^* & h_{14}^* & h_{24}^* & h_{34}^* & V_4-E & h_{0_{\rm II}
4}^* \\
0 & h_{0_{\rm II} 1}^{~} & h_{0_{\rm II} 2}^{~} & h_{0_{\rm II} 3}^{~} &
h_{0_{\rm II} 4}^{~} & - t e^{ik (E)} - E \\
\end {array}
\right]
\left[
\begin {array} {c}
B_{\rm I} \\
\psi_1 \\
\psi_2 \\
\psi_3 \\
\psi_4 \\
B_{\rm II} \\
\end {array}
\right]
=
\left[
\begin {array} {c}
\left( E + t e^{-ik(E)} \right) A_{\rm I} \\
- h_{0_{\rm I} 1}^* A_{\rm I} - h_{0_{\rm II} 1}^* A_{\rm II} \\
- h_{0_{\rm I} 2}^* A_{\rm I} - h_{0_{\rm II} 2}^* A_{\rm II} \\
- h_{0_{\rm I} 3}^* A_{\rm I} - h_{0_{\rm II} 3}^* A_{\rm II} \\
- h_{0_{\rm I} 4}^* A_{\rm I} - h_{0_{\rm II} 4}^* A_{\rm II} \\
\left( E + t e^{-ik(E)} \right) A_{\rm II} \\
\end {array}
\right],
\label {moleq}
\end {eqnarray}
\end {widetext}
where the components from top to bottom are respectively the FD equations
centered at sites $0_{\rm I}$, 1, 2, 3, 4, and $0_{\rm II}$. When $A_{\rm
I}$ and $A_{\rm II}$ are given, there are six unknowns ($B_{\rm I}$, 
$\psi_1$, $\psi_2$, $\psi_3$, $\psi_4$, and $B_{\rm II}$) to be found.
The unknowns can be found by a straightforward matrix inversion when the
square matrix has a nonzero determinant. Notably, the determinant actually
can {\em vanish} at certain energies for some system configurations, and
imply a nontrivial solution at $A_{\rm I} = A_{\rm II} = 0$ (no ingoing
waves). The solution must be localized within the molecule, since the
outgoing wave amplitudes $B_{\rm I}$ and $B_{\rm II}$ are necessarily
vanishing due to unitarity.

The determinant of the square matrix in Eq.~(\ref {moleq}) vanishes
whenever the rows or columns of the matrix are not linearly independent.
For instance, consider the configuration $h_{0_{\rm I} 2}^{~} = h_{0_{\rm
I} 3}^{~}$, $h_{0_{\rm II} 2}^{~} = h_{0_{\rm II} 3}^{~}$, $h_{12}^{~} =
h_{13}^{~}$, $h_{24}^{~} = h_{34}^{~}$, $h_{23}^{~} = h_{23}^*$, and $V_2
= V_3$, at the energy $E = V_2 - h_{23}$. In this occasion, the third and
fourth rows of the matrix are seen to be identical,
which means that the determinant of the matrix vanishes, and the solution
to the problem is not unique. Note that the system in this configuration
is not really ``symmetric'' in the usual sense.

A complete solution $\Psi (\lambda)$ for Eq.~(\ref {moleq}), in the case
of a simple symmetric system in which $h_{12} = h_{13} = h_{24} = h_{34}
\equiv \Delta$, $h_{0_{\rm I} 1}^{~} = h_{0_{\rm II} 4}^{~} \equiv
\Gamma$, and $h_{14}^{~} = h_{23}^{~} = h_{0_{\rm I} 2}^{~} = h_{0_{\rm I}
3}^{~} = h_{0_{\rm I} 4}^{~} = h_{0_{\rm II} 1}^{~} = h_{0_{\rm II} 2}^{~}
= h_{0_{\rm II} 3}^{~} = V_1 = V_2 = V_3 = V_4 = 0$ (which is similar to
the model considered in Ref.~\onlinecite {ZCP02}), at $E=0$ (i.e.,
$k=\pi/2$), where the determinant vanishes, is found to be 
\begin {widetext}
\begin {eqnarray}
\Psi (\lambda) &=& \psi_{\rm ext} (\lambda) + \beta \psi_{\rm loc}
(\lambda), \nonumber\\
\psi_{\rm ext} (\lambda) &=& { {t (A_{\rm I} - A_{\rm II})} \over {i
\Gamma} } (\Delta _{\lambda,1} - \Delta _{\lambda,4} ) - { {\Gamma} \over
{\Delta} } (A_{\rm I} \Delta _{\lambda,2} + A_{\rm II} \Delta_{\lambda,3}
) \nonumber\\
&& + \sum_{j=0}^\infty \left[ (i^{-j} A_{\rm I} + i^{j} A_{\rm II}) \Delta
_{\lambda, j_{\rm I}} + (i^{-j} A_{\rm II} + i^{j} A_{\rm I}) \Delta
_{\lambda, j_{\rm II}} \right], \nonumber\\
\psi_{\rm loc} (\lambda) &=& \Delta _{\lambda,2} - \Delta _{\lambda,3},
\label {psi0}
\end {eqnarray}
\end {widetext}
where $\beta$ is an arbitrary complex number, $i \equiv \sqrt {-1}$, and
$\Delta _{\lambda,\lambda'} = 1$ (0) when $\lambda = \lambda'$ ($\lambda
\neq \lambda'$). $\Psi$ is seen to be a superposition of an extended
state $\psi_{\rm ext}$ and a localized state $\psi_{\rm loc}$. When
$A_{\rm I} = A_{\rm II} = 0$ or $|\beta| \rightarrow \infty$, $\Psi
\rightarrow \psi_{\rm loc}$ and it is a localized state in the continuum
(LSC).

The origin of the LSCs in TB systems can also be understood in a more
direct manner, given an insight from the observation that $\psi_{\rm
loc}$ vanishes at the sites in the molecule in contact with the leads [see
Eq.~(\ref {psi0})].
In general, if $\psi_A$ and $\psi_B$ are respectively the stationary 
states at an energy $E$ in two isolated clusters of sites, labeled by $A$
and $B$, and $\psi_A^{~} (j_A^0) = \psi_B^{~} (j_B^0) = 0$, where $j_A^0$
($j_B^0$) is a site on cluster $A$ ($B$), then the direct product $\psi_A
\otimes \psi_B$ is a stationary state at $E$ in a system where clusters
$A$ and $B$ are coupled by $t_{AB}^{~} c^\dagger_{j_A^0} c^{~}_{j_B^0} +
{\rm H.c.}$ This is because though the FD equations for the coupled
clusters contain the additional terms $t_{AB}^{~} \psi (j_B^0)$ and 
$t_{AB}^* \psi (j_A^0)$, the wave function $\psi_A \otimes \psi_B$ is
still a solution of the FD equations since these terms vanish due to
$\psi_A^{~} (j_A^0) = \psi_B^{~} (j_B^0) = 0$. 
If one of the clusters, say cluster $A$, is infinitely large or open and 
$\psi_A$ is trivial, whereas cluster $B$ is finite sized and $\psi_B$ is
nontrivial, then $\psi_A \otimes \psi_B$ is a LSC. 
The generalization of the above argument to the case with more
than two clusters is straightforward. An example of such case is
the LSC in Eq.~(\ref {psi0}), which can be constructed from three
clusters, where two of them (the two leads) have infinitely large
sizes and trivial stationary states.

For $A_{\rm I} = 1$ and $A_{\rm II} = 0$, the transmission probability $T$
defined by $T \equiv |B_{\rm II}|^2$ is plotted versus the energy $E$ for
$\Delta = 0.2 t$ and $\Gamma = 0.4 t$ in Fig.~\ref {mol}(b). In the same
figure, $T$ is also plotted for the same system parameters but $V_2 =$
$0.05 t$ and  $0.3 t$, where the LSC is destroyed and has become an
almost-localized state. It is seen that $T$ can reflect the LSC only when
the LSC is destroyed by a perturbation and a Fano resonance appears. \cite
{VC06z} The blue shifts of the resonances from the energy of the LSC is
due to the increase of the energies of the almost-localized states by
$V_2$.

Therefore a comprehensive understanding of the problem may be stated as
the following. For problems of open systems, whenever the determinant of
the matrix to be inverted vanishes at an energy, there is a localized
state at that energy. If the energy is in a continuum of extended
scattering states, the localized state is a LSC, and a complete solution
is a superposition of the degenerate LSC and extended states. As the
states are decoupled, the transport which is related only to the extended
states does not reflect the presence of the LSC. When a LSC is destroyed,
or a previously localized state is coupled with the extended states, the
passing of a particle from one lead to the other through the molecule can
take place via two routes --- the extended states spanning the leads and 
the molecule or the almost-localized state in the molecule. That results
in a nonresonant and a resonant transmission amplitudes, and the
interference results in a Fano resonance. \cite {Fan35} When the
almost-localized state is on the verge of the decoupling from the
continuum and acquiring an infinitely long lifetime, the resonance is
infinitely sharp. \cite {VC06c}

\subsection {A quantum graph}
\label {qgsec}

The second example is a quantum graph. The quantum graphs are
multiply-connected 1D systems, which are meant to be effective models of
multiply-connected quasi-one-dimensional (Q1D) systems at low energies.
They are defined by the following conditions. Away from the junctions, a
particle is governed by a 1D Schr$\ddot {\rm o}$dinger equation. At a
junction, the wave functions on different branches are connected by a
chosen connecting scheme. \cite {VCT06} For a junction of three branches,
we have chosen a scheme defined by the equations (1) $\psi_1 = \psi_2 =
\psi_3$ and (2) $\nu \psi _1 + \sum_{i = 1} ^3 \partial \psi_i / \partial
x_i = 0$, where $\psi_i$ and $x_i$ are respectively the wave function and
coordinate defined on branch $i$. The coordinates are directed away from
the junction, and $\nu$ is a given real parameter with a dimension of
1/length.

\begin {figure}


\includegraphics{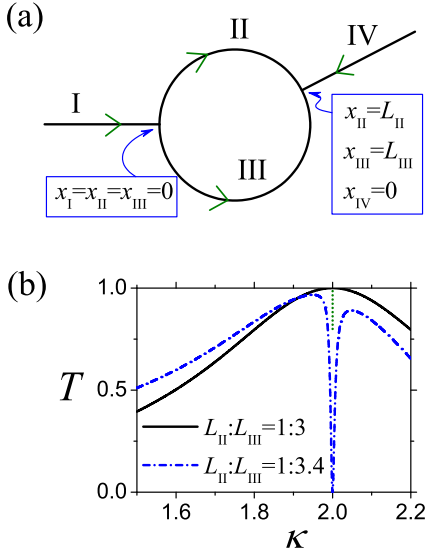}
\vspace {8cm}

\caption {
(Color online)
(a) The considered open quantum graph, which consists of a 1D ring (formed
by the branches labeled by II and III) connected to two leads (labeled by
I and IV). A coordinate $x_\eta$ with a positive direction indicated by an
arrow is defined on branch $\eta$ ($\eta =$ I, II, III, and IV).
(b) The transmission probability $T$ is plotted versus a dimensionless
wave number $\kappa$ defined by $\kappa \equiv k (L_{\rm II} + L_{\rm
III}) / (2\pi)$, for $\nu=0$, and $L_{\rm II} : L_{\rm III} = 1:3$ (solid
line) and $1:3.4$ (dash-dotted line).
There is a LSC for the $1:3$ case (indicated by a dotted line), but not
for the $1:3.4$ case in this energy range.
}

\label {qgraph}
\end {figure}

We consider a 1D ring connected to two 1D leads as shown in Fig.~\ref
{qgraph}(a). A potential everywhere equal to zero is assumed, and
the wave function at energy $E$ on the branch labeled by $\eta$ ($\eta = $
I, II, III, and IV) is $\psi ^\eta (x_\eta) = A_\eta e ^ {ikx_\eta} +
B_\eta e ^{-ikx_\eta}$, where $A_\eta$ and $B_\eta$ are arbitrary complex
numbers and $k \equiv \sqrt {2mE} / \hbar$. Applying the mentioned
connecting scheme at the two junctions, we obtain six simultaneous
equations or a matrix equation, 
\begin {widetext}
\begin {eqnarray}
\left[
\begin {array} {cccccc}
-1 & 1 & 1 & 0 & 0 & 0 \\
-1 & 0 & 0 & 1 & 1 & 0 \\
1-i\nu/k & 1 & -1 & 1 & -1 & 0 \\
0 & e^{ikL_{\rm II}} & e^{-ikL_{\rm II}} & 0 & 0 & -1 \\
0 & 0 & 0 & e^{ikL_{\rm III}} & e^{-ikL_{\rm III}} & -1 \\
0 & -e^{ikL_{\rm II}} & e^{-ikL_{\rm II}} & -e^{ikL_{\rm III}} &
e^{-ikL_{\rm III}} & 1-i\nu/k \\
\end {array}
\right]
\left[
\begin {array} {c}
B_{\rm I} \\
A_{\rm II} \\
B_{\rm II} \\
A_{\rm III} \\
B_{\rm III} \\
B_{\rm IV} \\
\end {array}
\right]
=
\left[
\begin {array} {c}
A_{\rm I} \\
A_{\rm I} \\
A_{\rm I} (1-i\nu/k) \\
A_{\rm IV} \\
A_{\rm IV} \\
A_{\rm IV} (1-i\nu/k) \\
\end {array}
\right],
\label {qgrapheq}
\end {eqnarray}
\end {widetext}
where $L_{\rm II}$ and $L_{\rm III}$ are respectively the lengths of
branches II and III. 
When the ingoing wave amplitudes $A_{\rm I}$ and $A_{\rm IV}$ are
specified, there are six unknowns to be found ($B_{\rm I}$, $A_{\rm II}$,
$B_{\rm II}$, $A_{\rm III}$, $B_{\rm III}$, and $B_{\rm IV}$).

From Eq.~(\ref {qgrapheq}), a LSC is seen at $k = k_n \equiv n n_0 \pi /
L_0$, when $L_{\rm II} : L_{\rm III}= n_{\rm II} : n_{\rm III}$, $n_0
\equiv {\rm min} (n_{\rm II}, n_{\rm III})$, $L_0 \equiv {\rm min} (L_{\rm
II}, L_{\rm III})$, $n$, $n_{\rm II}$, and $n_{\rm III}$ are
integers, and $n_{\rm II} + n_{\rm III}$ is even. The above condition
results in $e ^{ik_n L_{\rm II}} = e ^{ik_n L_{\rm III}} = (-1)^{n n_0}
\equiv \zeta_n$, and for the square matrix in Eq.~(\ref {qgrapheq}), the
differences between the corresponding elements in the first and second
rows are identical to that between the fourth and fifth rows. Therefore 
the rows are not linearly independent and the determinant of the matrix
vanishes. A complete solution $\Psi_n (x)$, where $x \in \{ x_\eta | ~
\eta = {\rm I}, {\rm II}, {\rm III}, {\rm IV} \}$, is found to be
\begin {widetext}
\begin {eqnarray}
\Psi_n (x) &=& \psi_n ^{\rm ext} (x) + \beta \psi_n ^{\rm loc} (x) ,
\nonumber\\
\psi_n ^{\rm ext} (x) &=& \sum _{\eta = {\rm I, IV}} \left[ A_{\eta} e ^
{i k_n x_{\eta}} + \Lambda_n ( \bar {A}_{\eta}, A_{\eta} ) e ^ {-i k_n
x_{\eta}} \right] \Delta_{x, x_{\eta}} \nonumber\\ 
&& + \sum _{\eta = {\rm II, III} } \left[ \Omega_n (A_{\rm I}, A_{\rm IV})
e ^{ik_n x_\eta} + \Omega_n (A_{\rm IV}, A_{\rm I}) e ^{ik_n (L_\eta -
x_\eta)} \right] \Delta_{x, x_\eta} , \nonumber\\
\psi_n ^{\rm loc} (x) &=& {\rm sin} ( k_n x_{\rm II} ) \Delta_{x, x_{\rm
II}} - {\rm sin} ( k_n x_{\rm III} ) \Delta_{x, x_{\rm III}}, \nonumber\\
\Lambda_n (X,Y) &\equiv& { { \zeta_n X - Y \nu_n } \over { 1 + \nu_n } },
~~~~~~ \Omega_n (X,Y) \equiv { { X (3 + \nu_n) + \zeta_n Y ( 1 - \nu_n ) }
\over  { 4 (1 + \nu_n) } } , \nonumber\\
\bar {A}_{\rm I} &\equiv& A_{\rm IV}, ~~~~~~  \bar {A}_{\rm IV} \equiv
A_{\rm I}, \nonumber\\
\nu_n &\equiv&  { \nu \over { i k_n} } ,
\label {qgsol}
\end {eqnarray}
\end {widetext}
where $\beta$ is an arbitrary complex number, and $\Delta_{x, x_\eta} =
1$ (0) when $x = x_\eta$ ($x \neq x_\eta$). The solution $\Psi_n$ is a 
superposition of a LSC $\psi_n ^{\rm loc}$ and an extended state
$\psi_n ^{\rm ext}$. Like in the case of a TB model discussed in
Sec.~\ref {tbmol}, $\psi_n ^{\rm loc}$ vanishes at the point in contact
with the leads

Since the LSC is decoupled from the extended states, it will not
be revealed in those spectral properties due to the scattering. For
quantum graphs, Texier \cite {Tex02} has pointed out that the Friedel sum
rule, which is related to the phases of the eigenvalues of the scattering
matrix, fails to count the number of states in a scattering region in
such a situation.
Experimentally, it has also been found \cite {SS92} that a 1D lead does
not couple to a 2D wave function when the lead is located at a node
of the wave functions.

The LSCs in the defined quantum graph can also be understood directly
from the connecting equations at the junctions. 
If a stationary state at energy $E$ in a graph labeled by $A$, has a node
at a point $P$ on one of the branches, then labeling the segments of the
branch on the two sides of $P$ by 1 and 2, defining coordinates $x_1$ and
$x_2$ respectively, with positive directions directed away from $P$, and
denoting the stationary wave functions on segments 1 and 2 by $\psi_1
(x_1)$ and $\psi_2 (x_2)$ respectively, one has $\psi_1 = \psi_2$ (both
are vanishing) and $\partial \psi_1 / \partial x_1 + \partial \psi_2 /
\partial x_2 = 0$ at $P$, and the stationary wave function can be written
in the form of the direct product $\psi_1^{~} \otimes \psi_2^{~} \otimes
\psi_{\rm others} ^A$, where $\psi _{\rm others} ^A$ is the direct
product of the wave functions on the other branches in the graph.
Similarly, for a graph labeled by $B$ containing a branch with an open end
labeled by 3, which has also a stationary state at $E$, the stationary
wave function can be written into the form $\psi_3^{~} \otimes \psi _{\rm
others} ^B$, where $\psi_3$ is the wave function on branch 3, and $\psi
_{\rm others} ^B$ is the direct product of the wave functions on the other
branches in the graph. 
If $\psi_3$ is trivial, when the open end of branch 3 is attached to $P$
by demanding the connecting equations $\psi_1 = \psi_2 = \psi_3$ and $\nu
\psi _1 + \sum_{i = 1} ^3 \partial \psi_i / \partial x_i = 0$ to be
fulfilled at $P$, the direct product $\psi_1^{~} \otimes \psi_2^{~} 
\otimes \psi_3^{~} \otimes \psi _{\rm others} ^A \otimes \psi _{\rm
others} ^B$ is a stationary solution at $E$ in the coupled graphs, since
the connecting equations are automatically fulfilled.
If graph $A$ is finite and $\psi_1^{~} \otimes \psi_2^{~} \otimes \psi
_{\rm others} ^A$ is nontrivial, whereas graph $B$ is infinitely large and
$\psi_3^{~} \otimes \psi _{\rm others} ^B$ is trivial, the stationary
state in the coupled graphs is a LSC.
Note that the above arguments can also be straightforwardly generalized to
the case of more than two clusters, and the LSC in Eq.~(\ref {qgsol}) is
an example of this.

For $A_{\rm I}=1$ and $A_{\rm IV}=0$, the transmission probability $T$
defined by $T=|B_{\rm IV}|^2$. In Fig.~\ref {qgraph}(b), $T$ is plotted
versus the wave number for $\nu = 0$, $L_{\rm II} : L_{\rm III} = 1:3$ and
$1 : 3.4$, at the vicinity of a LSC or almost-localized state. 
For an isolated ring with an uniform potential on it, the nodes of the 
stationary standing wave states are equally spaced, and hence 
``commensurate'' branch lengths (such as 1:3) in the open graph give rise
to LSCs. Note the Fano resonance when the LSC is destroyed.

\subsection {A two-dimensional waveguide}

\begin {figure}


\includegraphics{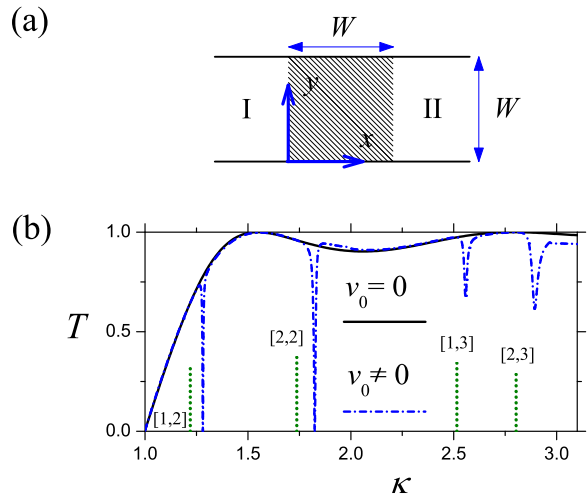}
\vspace {7cm}

\caption {
(Color online)
(a) The considered open 2D waveguide. Two leads (labeled by I and II) of
width $W$ are attached oppositely to a square cavity (shaded region). The
potential in the leads are kept at zero, and that in the cavity is kept at
$V_{\rm G}$. A coordinate system $(x,y)$ is defined in the cavity as
shown.
(b) The transmission probability $T$ is plotted versus a dimensionless
wave number $\kappa$ defined by $\kappa \equiv \sqrt {2mE} W / (\pi 
\hbar)$, for the case shown in (a) (solid line) and the case with an
additional $\delta$-potential $V(x,y) = v_0 \delta (x-W/6) \delta
(y-W/6)$ (dashed-dotted line).
In this energy range, LSCs (indicated by dotted lines) are found only in
the case of no $\delta$-potential. The pairs of integers in the form
$[m_x,m_y]$ indicate the profiles of the LSCs.
}

\label {waveguide}
\end {figure}

Our third example is a waveguide in a 2D continuous space as shown in
Fig.~\ref {waveguide}(a). The waveguide has a width of $W$, and the
potential in the waveguide is set at zero besides a square region that is
set at $V_{\rm G}$. We will call the infinitely extended regions on both
sides the leads, and the central finite sized region a resonating cavity.
This system may model a mesoscopic fabricated structure.

We discretize the continuous space into a square lattice, and the TISE
becomes a set of simultaneous FD equations that read \cite {Dat95,FG97} $-
t (\psi _{{\bf i}-\hat x} + \psi _{{\bf i}+\hat x} + \psi _{{\bf i}-\hat
y} + \psi _{{\bf i}+\hat y} ) + (V_{\bf i}-E+2t) \psi _{\bf i} = 0$, where
$E$ is the energy, $t \equiv \hbar^2/(2ma^2)$ ($a$ is the distance between
two nearest sites), $\hat x$ and $\hat y$ are respectively the unit
vectors along the $x$ and $y$ directions, ${\bf i} \equiv (i_x,i_y)$
(where $i_x$ and $i_y$ are integers), and $\psi_{\bf i}$ and $V_{\bf i}$
are respectively the wave function and potential at ${\bf i} a$. 
The FD equations are solved like in the first example.

In a lead labeled by $\eta$ ($\eta=$ I or II), taking the integer
$i_x^\eta$ ($i_y^\eta$) as the longitudinal (transverse) coordinate, the
wave function at energy $E$ for $V_{\bf i}=0$ is \cite {Dat95,FG97} 
\begin {eqnarray}
\psi ^\eta (i_x^\eta, i_y^\eta) &=& \sum_{m=1}^{N_ \eta} \left( A_m ^\eta
e ^{ik ^\eta _m i_x^\eta a} + B_m ^\eta e ^{-ik ^\eta _m i_x^\eta a}
\right) \nonumber\\ 
&& \times  {1 \over {\sqrt {N _\eta + 1}}} ~ {\rm sin} \left[ { {m\pi}
\over {N_\eta + 1} } (i_y^\eta+1) \right], ~~~
\end {eqnarray}
where $A_m ^\eta$ and $B_m ^\eta$ are arbitrary complex numbers, $i_y^\eta
= 0 \sim N_\eta-1$ [where $W \equiv (N_\eta + 1) a$], and $k^\eta _m$ is
defined by $E \equiv -2t \{ 2 - {\rm cos} (k ^\eta _m a ) - {\rm cos}
[{m\pi} / (N _\eta + 1)] \} $, where the cosine is defined by ${\cos } ~
\xi \equiv (e ^ {i\xi} + e ^{-i\xi} ) / 2$ for a complex $\xi$, and the
phase of $k_m ^\eta$ is chosen such that $A_m ^\eta$ is the amplitude of a
wave propagating or exponentially decaying inward.
The multiple transverse modes for the transverse coordinate $i_y^\eta$ is
a consequence of the quasi-one-dimensionality.

Taking the ingoing amplitudes $\{ A_m ^\eta \}$ as the input, the unknowns
will be $\{ B _m ^\eta \}$ (amplitudes of the waves propagating or
exponentially decaying outward) and the point-wise wave function in the
cavity $\psi _{\rm cavity}$. 
The FD equations centered at the sites within the cavity, and the
uniqueness requirement of the wave function at the interfaces between the
cavity and the leads results in a matrix equation of the form 
\begin {eqnarray}
M (E) \cdot | \{ B_m^\eta \}, \psi _{\rm cavity} \rangle = | \{ A_m^\eta
\} \rangle,
\label {2dguide}
\end {eqnarray}
where $| \{ A_m^\eta \} \rangle$ and $| \{ B_m^\eta \}, \psi _{\rm cavity}
\rangle$ are respectively the known and unknown column matrices, and
$M(E)$ is a square matrix whose determinant ${\rm det} M(E)$ may vanish
and imply LSCs in the system.

When there are no ingoing waves (i.e., $A_m ^\eta = 0$ for all $\eta$ and
$m$) and hence $| \{ A_m^\eta \} \rangle = 0$, a nontrivial solution for
$| \{ B_m^\eta \}, \psi _{\rm cavity} \rangle$ can be obtained if ${\rm
det} M(E) = 0$. This is necessarily a localized state since
the $B _m ^\eta$'s for the outward propagating waves necessarily vanish
due to unitarity, leaving only the possibility of nonvanishing $B _m
^\eta$'s for the outward exponentially decaying waves. If ${\rm det} M(E)
= 0$ occurs at an energy $E_0$ in the continuum, a complete solution at
$E_0$ is a superposition of the localized state found by ${\rm det} M(E) =
0$, and the extended states found by inverting $M(E)$ at $E_0+\delta$,
$\delta \rightarrow 0$.

Note that to find a nontrivial solution for the column vector $| \phi
\rangle$ in an equation $S | \phi \rangle = 0$, where $S$ is a square
matrix, is to find an eigenvector of $S$ that corresponds to a vanishing
eigenvalue (since the equation is just $S | \phi \rangle = {0 \cdot
| \phi \rangle }$). The eigenproblem can be solved by a numerical package
such as EISPACK (http://www.netlib.org/).
In general, there can be simultaneously more than one eigenvectors having
vanishing eigenvalues, and these eigenvectors are the degenerate localized
states in the original problem.

Figure \ref {waveguide}(b) shows the transmission probability $T$ for a
particle with an energy $E$, injected from the first subband in lead I,
and passed to the first subband in lead II. Letting $A_m^\eta = 
\Delta_{\eta,{\rm I}} \Delta_{m,1}$, $T$ is given by $T \equiv |B_1^{\rm
II}|^2$. 
We choose $V_{\rm G} = - 15 \hbar^2 / (m W^2)$, and consider the
cases with and without an additional perturbing $\delta$-potential $V
(x,y) = v_0 \delta (x-W/6) \delta (y-W/6)$ in the cavity, \cite {VC06h}
where $v_0 = 5 \hbar^2 / m$, and $x$ and $y$ are the coordinates defined
in Fig.~\ref {waveguide}(a).

In the considered energy range in Fig.~\ref {waveguide}(b), zeroes of
${\rm det} M(E)$ \cite {VC06a} or LSCs are found only for the case without
a perturbing $\delta$-potential. Note that LSCs can also exist at energies
beyond the first subband (at $\kappa > 2$).
We have used $N_\eta = 17$ in Fig.~\ref {waveguide}(b), and since we find
the LSCs qualitatively the same as those in the calculation using $N_\eta
= 8$, \cite {VC06} we believe they will survive in the continuous space
limit. 
As usual, Fano resonances appear when the LSCs are destroyed. 
The locations of the resonances depend on the details of the perturbing
potential, but when the perturbing potentials are vanishingly small, the
resonances are always found with infinitesimal widths on the locations of
the LSCs. A noteworthy point is the $\delta$-potential does not affect a
LSC when it is on a node of it.

\begin {figure}


\includegraphics{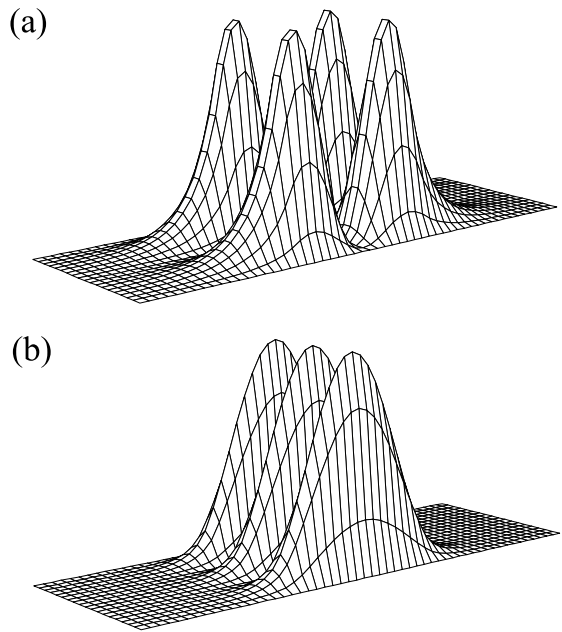}
\vspace {9cm}

\caption {
The probability densities of two of the LSCs in Fig.~\ref {waveguide}(b) 
are shown in a section of the waveguide with the cavity at the center. The
length of the section is three times of the width.
(a) The [2,2] LSC at $\kappa \simeq 1.74$ has sizable exponential tails in
the leads. Correspondingly, it has also a considerable red shift from
$\kappa _{\rm SW} (2,2) \simeq 2.23$, where $\kappa _{\rm SW} (m_x,m_y)$
is defined by $\kappa_{\rm SW} (m_x,m_y) \equiv \sqrt {2mE_{\rm SW}
(m_x,m_y)} W / (\pi \hbar)$.
(b) The [1,3] LSC at $\kappa \simeq 2.51$ has smaller exponential tails in
the leads, and also a lesser red shift from $\kappa_{\rm SW} (1,3) \simeq
2.64$.
}

\label {psi2}
\end {figure}

The LSCs in Fig.~\ref {waveguide}(b) have profiles with exponential tails
in the leads, and resemble the fictitious standing waves ${\rm sin} (m_x
\pi x / W) \cdot {\rm sin} (m_y \pi y / W)$ in the cavity, where $m_x$ and
$m_y$ are integers. Therefore we will use $[m_x,m_y]$ to label the LSCs
for the convenience in our discussion. The four LSCs in Fig.~\ref
{waveguide}(b), from the lowest to the highest energy, resemble standing
waves with $[m_x,m_y] = [1,2]$, [2,2], [1,3], and [2,3] respectively.
The probability densities or squares of the absolute values of the wave
functions of two of the LSCs are shown in Fig.~\ref {psi2}. The energy of
a LSC is found to be always red shifted from the energy of the
corresponding fictitious standing wave in the cavity $E_{\rm SW}
(m_x,m_y)$, where $E_{\rm SW} (m_x,m_y) = V_{\rm G} + (m_x^2 + m_y^2)
\pi^2 \hbar^2 / (2m W^2)$.
Moreover, it seems that the larger the exponential tails, the more the
red shift [comparing Figs.~\ref {psi2}(a) and \ref {psi2}(b)]. This is
conceivable since the tails lead to a lowering of the kinetic energies.


The presence of the above LSCs can be understood by an intuitive picture.
Notice that the first two LSCs which resemble the $[m_x,m_y]=$ [1,2] and
[2,2] standing waves, are embedded only in the first subband. Since their
transverse wave functions resemble ${\rm sin} (2 \pi y / W)$ and are 
orthogonal to the first transverse modes in the leads ${\rm sin} (\pi y /
W)$, the standing waves are trapped in the cavity.
With this picture, the absence of a LSC with $m_y=1$, such as a LSC with
$[m_x,m_y] = [2,1]$ or [3,1] is conceivable.
The two higher energy LSCs which resemble the standing waves with
$[m_x,m_y] = [1,3]$ and [2,3] are embedded in both the first and second
subband. Likewise, the trapping can be understood by the observation that
their transverse wave functions with $m_y=3$ are orthogonal to the
transverse modes in the leads with $m_y=1$ and 2. In this energy range we
do not find LSCs with $m_y=1$ and 2, such as [3,1] and [3,2]. \cite
{VC06f,SRW89,KSJ99,SBR06}

The LSCs as the eigenstates of $M(E)$ are orthogonal to each
other, though the $M(E)$ in Eq.~(\ref {2dguide}) is non-Hermitian in
general. The orthogonality can be argued from the fact that the LSCs are
localized and are not affected by a truncation of the leads at distances
far away from the cavity. Since the LSCs form a subset of the set of
eigenenergy states of the Hermitian Hamiltonian matrix for the truncated
(closed) system, they are orthogonal.

\section {Concluding Remarks}

The behaviors of the LSCs in open low-dimensional systems have been
illustrated by examples of various kinds. These LSCs are obtained by
studying the zeroes of the determinant of the matrix to be inverted in a
considered problem. A zero corresponds to at least one localized state.
The crucial factor in the formation of these LSCs is the
low-dimensionality of the leads, and the ``symmetricity'' in the systems
is not a necessary condition. \cite {VC06d}

For the TB molecule and quantum graph, the one-dimensionality of the leads
enables them to attach just at the nodes of a LSC in the scattering
region and thereby leaving the LSC intact.
The same argument also holds for the case of higher dimensional leads with
1D constrictions at the ends joining the scattering region.
For the case of Q1D leads, the delocalization of a standing wave in the
cavity can be prohibited by the non-overlapping of its transverse wave
function and the transverse modes in the leads, and the standing wave is
turned into a LSC. \cite {VC06b}

In view of the possibility of such LSCs not only in the case of idealized
1D leads but also in the case of Q1D leads, such LSCs may exist or may be
realizable in, e.g., mesoscopic structures \cite {GGH00} and nanobridges
or molecular junctions, \cite {SRG05} and may not of academic interest
only.

\indent {\bf Acknowledgments -} 
This work is supported by the National Science Council of Taiwan under
Grant No. 94-2112-M-009-017, and we also thank S.W. Chung for helpful
discussions.

\end{document}